\documentclass[12pt,english]{article}
\usepackage[T1]{fontenc}
\usepackage[latin9]{inputenc}
\usepackage{textcomp}
\usepackage{amstext}
\usepackage{graphicx}

\makeatletter
\newcommand{\pubnumber}{SNSN-323-63}
\newcommand{\pubdate}{\today}

\def\Title#1{\begin{center} {\Large #1 } \end{center}}
\def\Author#1{\begin{center}{ \sc #1} \end{center}}
\def\Address#1{\begin{center}{ \it #1} \end{center}}

\newcommand{\pubblock}{\rightline{\begin{tabular}{l} \pubnumber\\
         \pubdate  \end{tabular}}}
\newenvironment{Abstract}{\begin{quotation}  }{\end{quotation}}
\newenvironment{Presented}{\begin{quotation} \begin{center} 
             PRESENTED AT\end{center}\bigskip 
      \begin{center}\begin{large}}{\end{large}\end{center} \end{quotation}}

\def\beq{\begin{equation}}
\def\eeq#1{\label{#1}\end{equation}}
\def\eeqn{\end{equation}}

\def\beqa{\begin{eqnarray}}
\def\eeqa#1{\label{#1}\end{eqnarray}}
\def\eeqan{\end{eqnarray}}

\let\bar=\overbar

\def\Dslash{\not{\hbox{\kern-4pt $D$}}}
\def\dslash{\not{\hbox{\kern-2pt $\del$}}}

\def\msb{{\bar{\ssstyle M \kern -1pt S}}}

\makeatother

\usepackage{babel}

\begin{document}
\begin{titlepage} \pubblock

\vfill{}
 \Title{Measuring the angle $\gamma$ from penguin decays at LHCb} \vfill{}
 \Author{Angelo Carbone\\on behalf of the LHCb Collaboration} \Address{Istituto Nazionale di Fisica
Nucleare, Sezione di Bologna\\
Viale B. Pichat 6/2, 40127 Bologna, Italy } \vfill{}
 \begin{Abstract} In this paper we preset first LHCb
results on charmless charged two- and three-body B meson decays.
In particular, using the first pb$^{-1}$ of integrated luminosity,
LHCb observed the $B^{0}\rightarrow K^{+}\pi^{-}$
and $B^{+}\rightarrow K^{+}\pi^{+}\pi^{-}$ decays.
Such decays provide an interesing way to determine the angle $\gamma$
of the Unitary Triangle, thanks to the possible presence of New Physics
in the penguin loops. 

\end{Abstract} \vfill{}
 \begin{Presented} 6th International Workshop on the CKM Unitarity
Triangle\\
 Warwick, UK, Septermber 6--10, 2010 \end{Presented} \vfill{}
 \end{titlepage} \global\long\def\thefootnote{\fnsymbol{footnote}}
 \setcounter{footnote}{0}

\section{Introduction}

The family of charmless $B\rightarrow h^{+}h'^{-}$ , where $B$ can
be either a $ $$B^{0}$ or $B_{s}^{0}$ meson, while $h$ and $h'$
stand for, $K$ or $\pi$, are sensitive probes of the Cabibbo-Kobayashi-Maskawa
(CKM) \cite{CKM-1,CKM-2} matrix and have the potential to reveal
the presence of New Physics (NP) as they can occur via tree and penguin
amplitudes. One promising way to exploit the presence of penguins
for these decays was first suggested ten years ago in Ref. \cite{Fleischer-1}
(for the latest update of the analysis see Ref. \cite{Fleischer-2}).
In particular, it was shown how the combined measurement of the $B^{0}\rightarrow\pi^{+}\pi^{-}$
and $B_{s}^{0}\rightarrow K^{+}K^{-}$ $ $ time-dependent CP asymmetries,
under the assumption of invariance of the strong interaction dynamics
under the exchange of the $d\leftrightarrow s$ quarks (U-spin symmetry)
in the decay graphs of these modes, provides an interesting way to
determine the angle $\gamma$ of the Unitarity Triangle (UT), without
the need of any dynamical assumption. Due to the possible presence of NP in the penguin loops,
a measurement of $\gamma$ with these decays could differ appreciably from the one determined
by using other $B$ decays governed by pure tree amplitudes, e.g. $B\rightarrow DK$.

For what concerns the charmless charged three-body B meson decays, Ref. \cite{bediaga} depicts
a method to extract $\gamma$ by means of
Dalitz amplitude analyses of $B^{\pm}\rightarrow K^{\pm}\pi^{+}\pi^{-}$and
untagged $B^{0}$ and $\bar{B}^{0}\rightarrow K_{s}\,\pi^{+}\pi^{-}$.
The method is based on the ability to measure
independently the relative amplitudes and phases for $B^{0}$ and
$\bar{B}^{0}$ decays in a joint untagged sample. The following section 
will report the first experiences of reconstructing these decays with the first pb$^{-1}$
of integrated luminosity collected by LHCb.

LHCb \cite{LHCb} is a dedicated
B physics experiment which exploits the unprecedented quantity
of $b$ hadrons produced at the LHC to over-constrain the CKM matrix
and search for New Physics (NP) in the flavour sector. The
$b\bar{b}$ cross section at the center of mass energy of 7 TeV
was first measured by LHCb as $\sigma_{b\bar{b}}=(75.3\pm5.4\:(stat.)\pm13.0\:(syst.))$ mb
in the pseudorapidity interval $2<\eta<6$ \cite{BEAUTYXSECT}.

\section{Charmless charged two and three-body B meson decays}
\subsection{The trigger}

The LHCb trigger is a two level system consisting of Level 0 (L0) and High Level
Trigger (HLT). The L0 is implemented in hardware, it is designed
to reduce the visible interaction rate to a maximum of $1$ MHz, at
which the whole detector can be readout. 

The High Level Trigger (HLT) is the second (and last) level of trigger
of LHCb, running on events passing the L0 trigger. It consists of
a C++ application that runs on every CPU of the Event Filter Farm
(EFF), which will be made out of 1000 16-core computing nodes when 
it will fully operate in 2011. The HLT application has access to all data in one event.

The charmless charged two- and three-body B decays
are triggered by the L0 requiring at least one cluster in the hadronic calorimeter
with a transverse energy higher then a threshold that depends on
the rate of the proton-proton interaction. Then the HLT, using a
fast track reconstruction, applies a cut on transverse momentum and
impact parameter with respect to the primary vertex on the track associated
to the triggered L0 cluster. 

\subsection{The offline selection}

The charmless charged two- and three-body B meson decays, due to their
branching ratios in the range of $10^{-5}-10^{-6}$, are selected
by applying a set of offline selection cuts necessary to reduce the huge amount
of combinatorial background. Inclusive selection strategies are used for both the
two and three body decays studies. These are based on a two or three
charged tracks selection without particle identification, assigning
all tracks the pion mass hypothesis. The discriminating variables
include transverse momentum and impact parameters of the daughter tracks, with in addition the
common vertex $\chi^2$ and the distance of flight of the B meson candidate. The final analysis makes
use of the excellent
particle identification provided by the LHCb RICH system. The clear
separation between pions, kaons and protons allows very clean invariant
mass signals to be obtained for each decay.

\subsection{$B\rightarrow h^{+}h'^{-}$ decays: results and prospects}

Amongst the various modes, the decay with the highest
branching ratio, i.e. the $B^{0}\rightarrow K^{+}\pi^{-}$ ($\mathcal{BR}(B^{0}\rightarrow K^{+}\pi^{-})=(19.4\pm0.6)\times10^{-5}$$ $
\cite{PDG}), has been clearly observed. 

%%%%%%%%%%%%%%%%%%%%%%%%%%%%%%%%%%%%%%%%%%%%%%%%%%%%%%%%%%%%%%%%%%%%%%%%%
%%
%%   use this format to include an .eps figure into your paper
%%
%
\begin{figure}[htb]
\centering \includegraphics[scale=0.2]{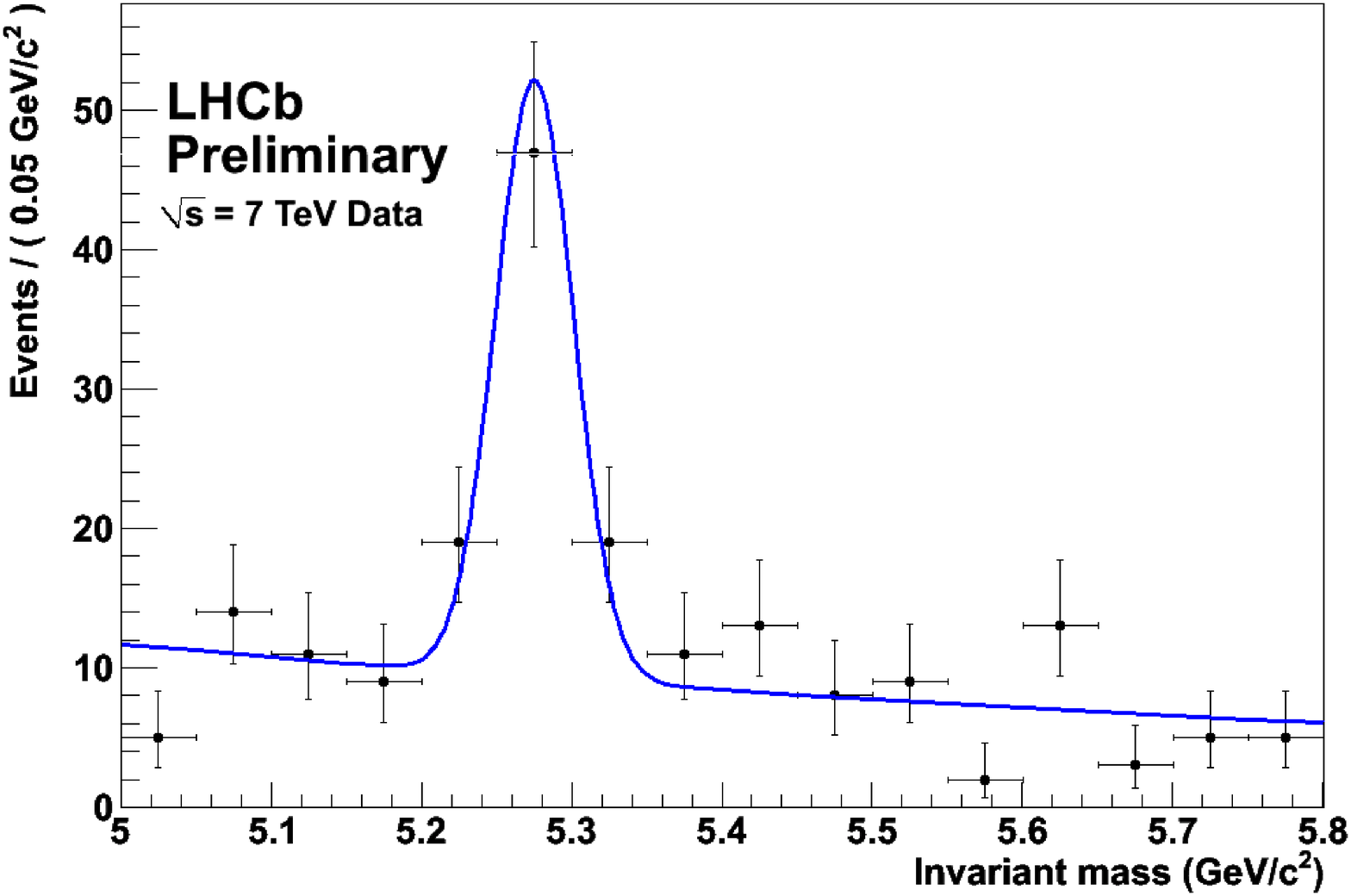}\includegraphics[scale=0.18]{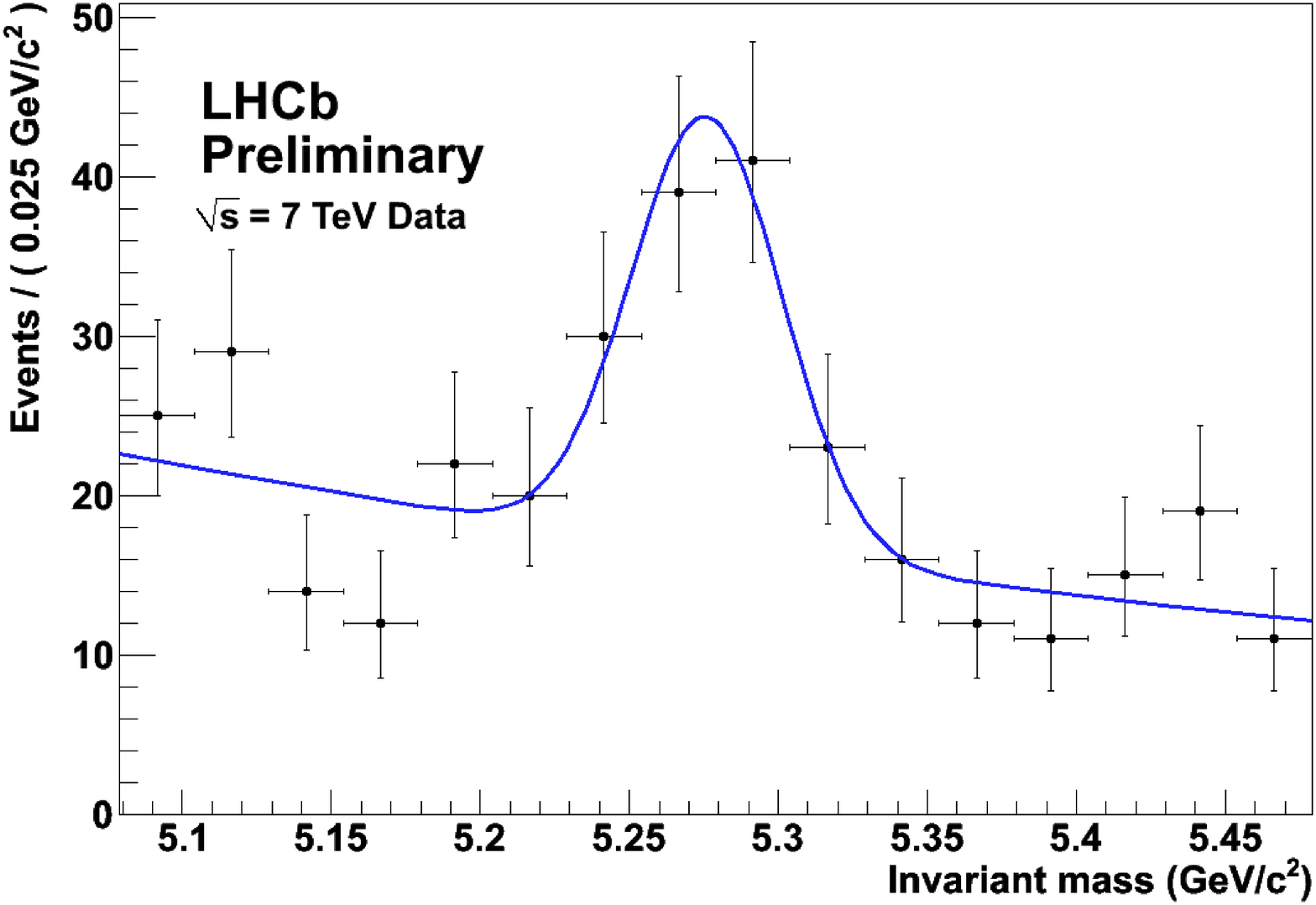}
\caption{$B^{0}\rightarrow K^{+}\pi^{-}$ invariant mass peak obtained by analyzing
$0.9$ pb$^{-1}$ of integrated luminosity (left). $B^{+}\rightarrow K^{+}\pi^{+}\pi^{-}$
invariant mass peak obtained by analyzing $0.55$ pb$^{-1}$ of integrated
luminosity (right).}

\label{fig:magnet} 
\end{figure}

%%%%%%%%%%%%%%%%%%%%%%%%%%%%%%%%%%%%%%%%%%%%%%%%%%%%%%%%%%%%%%%%%%%%%%%%%%%

The invariant mass peak of the $B^{0}\rightarrow K^{+}\pi^{-}$ obtained
by analyzing $0.9$ pb$^{-1}$ of integrated luminosity is shown in
Figure~\ref{fig:magnet} (left). The number of signal events is $N_{B^{0}\rightarrow K^{+}\pi^{-}}=56\pm10$,
while the direct $CP$ asymmetry, defined as $A_{K\pi}^{CP}=\frac{N_{\pi^{+}K^{-}}-N_{\pi^{-}K^{+}}}{N_{\pi^{+}K^{-}}+N_{\pi^{-}K^{+}}}$, is found to be $A_{K\pi}^{CP}=0.01\pm0.16$, where only the statistical
error is quoted. This is consistent with the world
average ($A_{K\pi}^{CP}=-0.098\pm0.012$ \cite{HFAG}). %In addition also an evidence of the decay  ( (ref. pdg)) has been observed with a yield of  .

Detailed studies about the prospects for measuring $CP$ violation,
relative branching ratios and the $\gamma$ angle from
$B\rightarrow h^{+}h'^{-}$ at LHCb can be found in the LHCb Roadmap
document \cite{roadmap}. At that time a $b\bar{b}$ cross section $\sigma_{b\bar{b}}\sim500\:\mu$b at $14$
TeV was assumed. Even with the reduced $b\bar{b}$
cross section measured at $7$ TeV, LHCb will be able to perform
competitive measurements already using a few tens of
pb$^{-1}$ of integrated luminosity, in particular to measure the
direct CP asymmetry, the relative branching ratio of the $B_{s}^{0}\rightarrow K^{-}\pi^{+}$
decay and the $B_{s}^{0}\rightarrow K^{+}K^{-}$ lifetime. In order
to overcome the current world statistics and measure time
dependent CP asymmetries, it will be necessary to accumulate about $500$
pb$^{-1}$ of integrated luminosity. 

For what concerns the angle $\gamma$ measurement, according to the
Monte Carlo studies based on proton-proton collisions
at $14$ TeV, it is expected to have a sensitivity
on $\gamma$ of about $7^{\text{\textdegree}}$
with an integrated luminosity of $2$ fb$^{-1}$.

\subsection{$B \rightarrow hhh$: results and prospects}

The observation of the decay $B^{+}\rightarrow K^{+}\pi^{+}\pi^{-}$,
($\mathcal{BR}(B^{+}\rightarrow K^{+}\pi^{+}\pi^{-})=(5.1\pm0.29)\times10^{-5}$ \cite{PDG}),
has been established by analyzing a sample of $0.55$ pb$^{-1}$ of
integrated luminosity. The invariant mass peak is shown in Figure~\ref{fig:magnet} (right).
The yield is $69\pm14\mbox{}$. Extrapolating this result to an integrated
luminosity of $1$ fb$^{-1}$, LHCb will be able to collect about
$100$k $B^{+}\rightarrow K^{+}\pi^{+}\pi^{-}$, $30$k $B^{+}\rightarrow\pi^{+}\pi^{+}\pi^{-}$,
$10$k $B^{+}\rightarrow K^{+}K^{-}\pi^{-}$, $60$k $B^{+}\rightarrow K^{+}K^{+}K^{-}$,
$3$k $B^{+}\rightarrow pp\pi^{-}$ and $10$k $B^{+}\rightarrow ppK^{-}$,
assuming that all the modes have the same reconstruction and trigger
efficiency. LHCb will so surpass
the current world statistics by one order of magnitude. 

The resonance structures present in three body charmless $B^{+}$ decays
can be a rich field for CP studies. A model independent approach will
be employed to search for statistically significant Dalitz differences
between the charged conjugate three body decays. The method is suggested
in the Ref. \cite{THEMETHOD}. %The so called Mirandizing approach ref takes each Dalitz surface, divides it into i bins,  for  Dalitz and  for  and computes the CP bin significance distribution . , which gives the amount of standard deviation assuming a Poissonian distribution. The  distribution must be a Gaussian centred on zero and of width 1 , for two statistically equivalent Dalitz surfaces, which would be the case for a CP conserving decays. Otherwise, in the presence of a CP violating source, the  distribution will be different from a .

With the large statistics samples we expect at LHCb in 2011,
it will be possible to subdivide the Dalitz plot
into different regions to identify what resonance or resonances are
the origin of the CP violation. For the decays which have branching
fractions bigger than $10^{-5}$ ($B^{+}\rightarrow K^{+}\pi^{+}\pi^{-}$,
$B^{+}\rightarrow\pi^{+}\pi^{+}\pi^{-}$ and $B^{+}\rightarrow K^{+}K^{+}K^{-}$)
it is expected to study possible sources of CP violation, while for
the other decays the sensitivity will be lower.

\section{Conclusions}

The analysis of the first pb$^{-1}$ of integrated luminosity collected
by LHCb shows very encouraging results for
the charmless charged two- and three-body B meson decays.
We have already observed the $B^{0}\rightarrow K^{+}\pi^{-}$
($N_{B^{0}\rightarrow K^{+}\pi^{-}}=56\pm10$) and $B^{+}\rightarrow K^{+}\pi^{+}\pi^{-}$
($N_{B\rightarrow K^{+}\pi^{+}\pi^{-}}=69\pm14$) decays, with yields in agreement with expectations.
Assuming an integrated luminosity of over $0.5$ fb$^{-1}$, 
LHCb will be competitive with and even overcome current
world statistics at some point in 2011.

\end{document}